\documentclass[aps,prl,twocolumn,groupedaddress]{revtex4}
\usepackage{graphicx}
\usepackage{dcolumn}
\usepackage{bm}
\begin{document}

\title{Spin transference and magnetoresistance amplification in a transistor}
\author{H.~Dery}\email{hdery@ucsd.edu}
\author{\L.~Cywi{\'n}ski}
\author{L.~J.~Sham}
\affiliation{Department of Physics, University of California San
Diego, La Jolla, California, 92093-0319}
\date{\today}

\begin{abstract}
A current problem in semiconductor spin-based electronics is the
difficulty of experimentally expressing the effect of spin-polarized
current in electrical circuit measurements. We present a theoretical
solution with the principle of transference of the spin diffusion
effects in the semiconductor channel of a system with three magnetic
terminals. A notable result of technological consequences is the
room temperature amplification of the magneto-resistive effect,
integrable with electronics circuits, demonstrated by computation of
current dependence on magnetization configuration in such a system
with currently achievable parameters.
\end{abstract}
\maketitle

Giant magnetoresistance effect has been discovered in
heterostructures of ferromagnetic and paramagnetic metal layers
\cite{Baibich_PRL88,Binasch_PRB89}. Similar effect has been observed
in magnetic tunnel junctions \cite{Moodera_PRL95}. Applications in
the spin valve configuration has given rise to important products
such as hard disk read-heads and magnetic memories
\cite{Prinz_Science98}. Research into spin polarized currents in
semiconductors leads to the new field of semiconductor spintronics
\cite{Wolf_Science00,Zutic_RMP04}  with promise of increased logic
functionality of electronic circuits and integration with
non-volatile magnetic memory. While room temperature injection from
a ferromagnet via a tunnel barrier into a semiconductor has produced
reasonable current spin polarization
\cite{Hanbicki_APL02,Hanbicki_APL03,Jiang_PRL05,Adelmann_PRB05}, the
magneto-resistive effect in a semiconductor spin valve with
ferromagnetic metal contacts is predicted to be small
\cite{Fert_PRB01,Rashba_EPJB02}.
In order to introduce
additional control over the spin-polarized carrier flow, a number of
semiconductor-based spin-transistors have been proposed
\cite{DattaDas_APL90,Schliemann_PRL03,Ciuti_APL02,Zutic_PRL02,Flatte_APL03},
using ideas such as Rashba effect, effective spin reflection,
half-metallic ferromagnets,  or minority carrier action in junctions
with a magnetic semiconductor.

In this Letter we present an electrical means of expressing the spin
effects, rather than optical means
\cite{Crooker_PRL05,Stephens_PRL04}. We study the diffusive spin
currents of a nonmagnetic semiconductor (SC) with three
ferromagnetic metal (FM) terminals, each capable of injecting or
extracting spin-polarized currents. The currents flowing between the
contacts depend on the alignment of their magnetizations. The
magneto-resistive (MR) effect is defined as a relative change of the
current upon flipping of one of the magnetization vectors. In the
diffusive regime considered here, the MR effect comes from spin
accumulation in the semiconductor due to the spin selectivity of the
contacts. The profiles of non-equilibrium spin densities in the
semiconductor depend on the magnetic configuration, and the
resulting different diffusion currents are the cause of MR. Using
more than two terminals provides the capability of transference of
spin effects from one controlled diffusion region to another. This
transference under voltage and magnetization control will be shown
to lead to amplification of the magneto-resistive effect. This
control provided by a third conducting and biased terminal is in
contrast to the ``nonlocal'' spin valve geometry
\cite{Jedema_Nature01,Johnson_Science93}, where the additional
contact is a floating voltage probe.
\begin{figure}
\includegraphics[height=4cm,width=8.6cm]{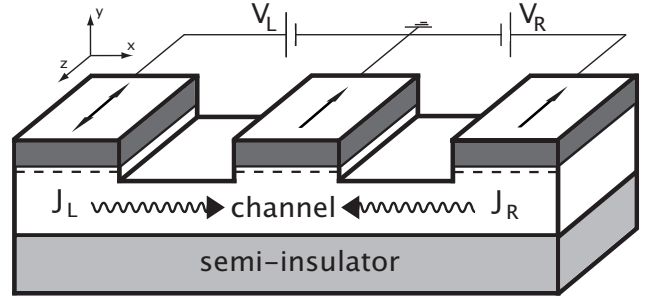}
\caption{A schematics of the proposed three-terminal device.  The
channel indicates the current flow region of $n$-doped semiconductor
grown on top of an insulating substrate.} \label{fig:scheme}
\end{figure}

An illustrative application of these ideas is a device we term
magnetic contact transistor (MCT), shown in Fig~\ref{fig:scheme}.
The principle behind such a system is analogous to the familiar
bipolar transistor. Here the two spin diffusion channels whose
populations can change by spin-flip take the place of electrons and
holes whose populations are changed by recombination. In the latter
case, the transistor action results from attaching two p-n diodes
back to back into a pnp or npn structure, in which the common base
width must be smaller than the recombination diffusion length. A
longer base would beget two uncoupled diodes devoid of amplification
effects. The two-terminal device in spintronics is a spin-valve, in
which the current depends on the relative magnetization directions
of two magnetic contacts. By connecting two spin valves with a
common source terminal (the middle contact in
Fig.~\ref{fig:scheme}), whose width is smaller than spin-diffusion
length $L_{sc}$, we create a system capable of amplifying the MR
effect of each spin valve. In contrast to the bipolar transistor,
the spin current is driven by spin diffusion rather than charge
diffusion. In addition to the current control by applied voltages as
in the conventional transistor, this spin transistor has control of
the spin components by the magnetization configurations of the
ferromagnetic electrodes. One strategy in amplifying the MR effect
is  based on the ability to use the voltage adjustment to yield  a
zero or near zero current in one magnetic configuration of the
ferromagnetic electrodes, and the ability to produce an easily
measurable current at the same voltage arrangement by changing the
magnetic configuration. The former is simple circuit theory and
requires no spin physics but the latter is a result of intricate
spin transport physics in the transistor configuration. In our
scheme, it requires the coupling of three contacts via spin
currents, equivalent to coupling spin transport of two spin valves.
The ratio of the ``on'' current to the ``off'' current will be shown
below to be robust against electrical noise and to decay with
increasing width of the middle contact or, equivalently, the
distance between two spin valves beyond the spin diffusion length,
indicative of the essential role of  spin.

The system shown in Fig.~\ref{fig:scheme} is a planar structure. Low
mesas beneath the FM terminals are heavily doped, making the
Schottky barriers thin ($<$10 nm). As an illustration here, we let
the left drain (L) be a ``soft'' magnetic layer whose magnetization
can be easily flipped. The middle source (S) and the right drain
(R), are magnetized in the same direction. The P and AP
configurations denote, respectively, the L magnetization parallel
and antiparallel to that of the S and R. The currents $J_{L(R)}$ are
flowing in the L (R) part of the channel, and are measured in the
L(R) contacts, which are kept at separately controlled voltages
$V_{L(R)}$. The required magnetic properties of contacts can be
achieved either by pinning the magnetization of middle and right
terminals, or by exploiting the contacts' magnetic shape anisotropy,
in order to manipulate their coercivities. The magnetization of the
left drain can be controlled by external magnetic field (then our
system works as a sensor of $B$ field), or by a field created by a
pickup current in a wire above the contact. Analogously to CMOS
transistors, the MCT has the capacity for digital operation that can
be used to trigger a pickup current of another MCT, thus
transferring the information from one magnet to another, leading to
a new paradigm of computation \cite{Prinz_Science98,Ney_Nature03}.

We assume the system to be homogeneous in the $z$ direction (see
Fig.~\ref{fig:scheme}) and consider  two-dimensional spin diffusion
in the semiconductor:
\begin{equation}
\nabla^2 \mu_{s}(x,y) = \frac{\mu_{s}(x,y) -
\mu_{-s}(x,y)}{2L_{sc}^2} \,\, , \label{eq:diffusion}
\end{equation}
where $\mu_{s}$ is the spin dependent electrochemical potential with
$s$=$\pm$ denoting the spin species, and the spin diffusion length
$L_{sc}=\sqrt{D\tau_{sp}}$, where $D$ is the diffusion constant and
$\tau_{sp}$ is the spin relaxation time. This equation is well known
for paramagnetic metals \cite{Hershfield_PRB97}, and holds for
non-degenerate semiconductors considered here when the electric
field is small \cite{Yu_Flatte_long_PRB02}. This condition is
fulfilled here because of the presence of highly resistive barriers
at the FM/SC interface. In the following calculations  the electric
driving force on the current is shown to have negligible influence
on spin diffusion. Due to vastly different resistances in metals and
semiconductors, we can neglect the spin and spatial dependence of
the electrochemical potential in the ferromagnets and replace
$\mu_{s}^{FM}$ by a constant given by the bias voltage. Thus, we
only need to solve the diffusion equation inside the semiconductor
channel. We neglect interfacial spin scattering and use Ohm's law
across the contacts, so that the boundary conditions are:
\begin{eqnarray}
ej^i_s=\frac{\sigma_{sc}}{2}\Big(\widehat{n}\cdot\nabla\mu_{s}\Big)
\!=\! \left\{\!\!\!
\begin{array}{ll} -G^i_s (eV^i + \mu_s), &  \text{contacts}   \\ \\ \qquad 0, &
\text{otherwise},
\end{array} \right. \label{eq:boundaries}
\end{eqnarray}
where $j^i_s$ denotes the spin $s$ current through the $i^{th}$
barrier interface, $\sigma_{sc}$ the semiconductor conductivity,
$\widehat{n}$ the outward interface normal and $G^i_{s}$ the
spin-dependent barrier conductance. The spin-selective properties of
the tunneling barrier ($G_+$$\neq$$G_-$) dominate the spin injection
physics \cite{Rashba_PRB00}. This approach is valid for contacts
without depletion beyond the thin doped tunneling barrier
\cite{Albrecht_PRB03}. Elsewhere we will derive from the 2D
diffusion an effective 1D formalism for the lateral spin transport
inside a thin layer, with finite-sized metal contacts on top of it.
The calculations presented below were done with both methods giving
essentially the same results. Confining the spatial extent of spin
accumulation to the terminals' footprint is optimal for MR
amplification. It is achieved by etching away the semiconductor
peripheries in Fig.~\ref{fig:scheme}, or by selectively doping the
channel with ion implantation.

The barrier conductances $G_{s}$ have been calculated for Fe/GaAs
system in the simple single-band effective mass model
\cite{tunneling_phenomena}. We have assumed triangular Schottky
barriers of $\sim$7 nm thickness. At low applied voltages we obtain
the conductances of the order of $10^{2}$
($10^{3}$)$\,\Omega^{-1}$cm$^{-2}$ for reverse (forward) bias. The
ratio of reverse to forward bias conductance $f$ is taken to be 0.5,
a value poorer than the theoretical estimate of 0.2. The ratio of
spin-up to spin-down conductance is $G_{+}/G_{-}\simeq 2$. The
corresponding spin-injection efficiency coefficient
$\alpha=(G_{+}-G_{-})/(G_{+}+G_{-})$ is $30$\% as seen in circular
polarization degree of spin light emitting diodes
\cite{Hanbicki_APL02,Hanbicki_APL03}.
The spin selectivity of the
interface is governed by the ratio of Fermi velocities for up and
down spins in Fe. While this simple model lacks details of the
atomic structure of the interface \cite{Butler_JAP97}, both the
$G_{+}/G_{-}$ ratio and the orders of magnitude of $G_{s}$ are in
agreement with the experimental results.

It is instructive to look at the behavior of one half of the
semiconductor channel decoupled from the other as a two-terminal
spin valve. We define $MR$$=$$(J^P-J^{AP})/J^{P}$, with $J^{P}$
($J^{AP}$) denoting the total current through the structure for
parallel (antiparallel) alignment of magnetizations of the two
terminals. We consider a GaAs layer at room temperature with a
conservative value of $L_{sc}$$=$$1~\mu$m. It corresponds to a
spin-independent mobility of $5000$ cm$^{2}/$(Vs) and spin
relaxation time $\tau_{s}$$=$$80$ ps in the non-degenerate regime at
room temperature \cite{Optical_Orientation}. The free carrier
concentration is $n$$=$$4\cdot10^{15}$cm$^{-3}$. The lower spin
conductance of the forward-biased barrier
$G_{b}=1000$$\,\Omega^{-1}$cm$^{-2}$.
The dimensions are the following: the contact width $w$ and the
separation between the contacts $d$ are both 200 nm, and the
thickness of the semiconductor channel is 100 nm. For these
parameters we obtain the spin-valve MR$\approx$3\%. Such a small
effect is hard to measure and useless for applications. The weak
effect can be understood from a simplified 1D transport picture
where the analytical solutions are easily available
\cite{Fert_PRB01,Rashba_EPJB02}. Although such analysis ignores the
lateral geometry issues it gives a qualitative description within an
order of magnitude estimate. Even for Schottky barriers as thin as
the ones considered here, and for non-degenerate semiconductors, the
effective conductance $G_{sc}$$=$$\sigma_{sc}$$/$$L_{sc}$ is much
higher than $G_{b}$. The calculation gives, to the lowest order in
$G_{b}/G_{sc}$, $MR\sim \alpha^2 \frac{G_{b}}{G_{sc}}
\frac{L_{sc}}{d} \frac{f}{f+1}$. In the two-terminal case the
difference between P and AP configurations is accommodated by
different non-equilibrium spin-density profiles, with a very small
change in the total current. Although $J^{P}\simeq J^{AP}$, the
electrochemical potential splitting $\Delta \mu =
|\mu_{+}^{SC}-\mu_{-}^{SC}|$ near the contact follows the ratio:
\begin{equation}
\frac{\Delta \mu^{P}}{\Delta \mu^{AP} } \propto
\Big(\frac{d}{2L_{sc}}\Big)^{2} \ll 1 \label{eq:delta_mu},
\end{equation}
whereas the mean value of $\mu_s$ does not change visibly between P
and AP. Our three terminal scheme makes effective use of
Eq.~(\ref{eq:delta_mu}).

\begin{figure}
\includegraphics[height=5cm,width=8.6cm]{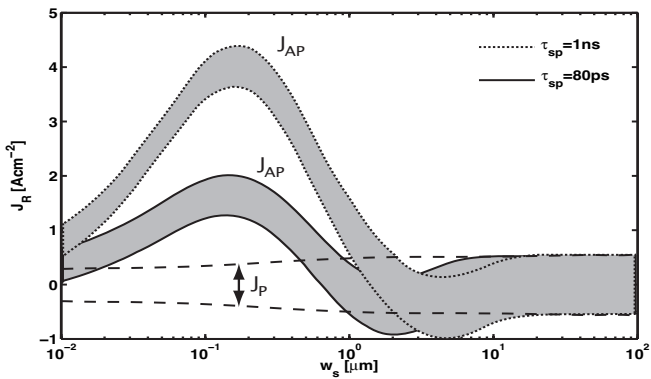}
\caption{ \footnotesize{Right drain current density versus source
width for the left drain magnetization parallel  (P) and
antiparallel (AP) to the other two magnets. $V_L=0.1V$ and $V_R$ is
adjusted for each $w_{s}$ so that $J_R^{P}=0$.  Upper and lower
curves of each set show the current densities for $V_R\pm$0.2~mV.
The dashed lines are the values for $J^{P}_R$ around zero current.
The solid and dotted lines show the currents for  $J^{AP}_R$ for two
different spin-flip times. The drain width, separation between the
contacts and the channel thickness are
\protect{$w$=$d$=$2h$=$0.2~\mu$m}. }}\label{fig:transistor_effect}
\end{figure}

For a certain  ratio $V_R/V_L$, the right contact current in the P
configuration can be quenched ($J_{R}^{P}=0$), while the AP current
remains finite. Thus, the readout of the magnetic configuration is
digitized. From simple circuit theory, when the barrier resistances
dominate, the voltage ratio $V_R/V_L$ which quenches $J_{R}^{P}$ is
estimated to be $1/(rf+1)$, where $r$ is the ratio of source to
drain widths. We examine the tolerance to error or noise by varying
this ratio by $\pm 0.2\%$, consistent with the Johnson noise for the
barrier resistance with contact area of 1~$\mu$m$^2$ at sub-GHz
frequencies. Fig.~\ref{fig:transistor_effect} shows the noise
margins for the right terminal current as function of the width of
the source contact $w_{s}$. For large source width, equivalent to
uncoupled spin valves, the MR amplification effect is lost. For
ultra-narrow contacts the effect is also small due to the increased
resistance of the contact \cite{Fert_PRB01}. In-between there is  an
optimal value of $w_{s}$ (well below $L_{sc}$)  where the MR effect
measured in the right contact easily exceeds hundreds of percent.
The resulting current densities calculated at room temperature are
of the order of 1 A/cm$^{2}$, which could be directly measured in
sub-micron contacts or further amplified, leading to a robust
read-out of the left drain magnetization direction.
During the read-out the left contact current density is about 100
A/cm$^{2}$ for the voltages used in
Fig.~\ref{fig:transistor_effect}.

\begin{figure}
\includegraphics[height=6cm,width=8.6cm]{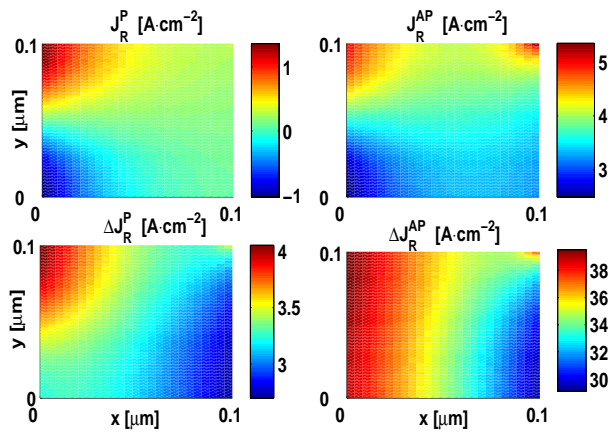}
\caption{ \footnotesize{ (color). Spatially resolved $x$ component
of the current density at the right half of the right channel. Note
the different scales for each figure. The upper panels show the zero
and finite net charge currents for the parallel and antiparallel
configurations, respectively ($0$ versus $\sim$$3.5$Acm$^{-2}$). The
lower panels show the amplification of $\Delta J$$=$$j_+$$-$$j_-$
due to the difference in electrochemical potential splitting. The
parameters are as in Fig.~\ref{fig:transistor_effect} with optimal
source width and $\tau_{sp}$$=$$80$ ps. }}\label{fig:2d_resolved}
\end{figure}

The amplification of the MR effect depends on the robustness of a
finite current $J_{R}^{AP}$  when the magnetic configuration is
changed to AP. An explanation based on our calculation is as
follows. The source of the different currents is the difference in
the spin splitting of the electrochemical potential, $\Delta$$\mu$,
at the R contact in the P and AP configurations at the optimal
voltage ratio. In P, $\Delta$$\mu^{P}$ is small. In AP,
$\Delta$$\mu^{AP}$ is large, for the reason in
Eq.~(\ref{eq:delta_mu}), which still qualitatively holds in the MCT.
This disrupts the balance of the spin
currents on two sides of the source, resulting in a sizable current,
$J_R^{{AP}} \sim (G_{+}-G_{-})\Delta \mu^{{AP}}$. In the
semiconductor channel, the electrochemical potential of one spin
population rises a good fraction above $-eV_R$ and the other one
drops an equal amount below. Thus,  the current is fully
spin-polarized. Fig.~\ref{fig:2d_resolved}  shows the two dimensional
spatial distribution of the $x$-component current densities in the
semiconductor channel in the $100$~nm section left of the right
contact. The middle contact width is at the optimal value shown in
Fig.~\ref{fig:transistor_effect}.
Note that the difference in total current densities between
Figures~\ref{fig:transistor_effect} and \ref{fig:2d_resolved} comes
from the ratio of channel thickness $h$ and R contact width $w$. In
Fig.~\ref{fig:2d_resolved}, the zero and 4 A/cm$^2$ respective
values in the middle of the color scale for the left and the right
upper panels show the zero current in P and a robust finite current
in AP. The amplification of the spin current from P to AP is visible
in the lower panels. By adding and subtracting the current densities
in the upper and lower panels of each column,  one can see that the
spin currents in the P case flow in the opposite directions at the
opposite edges of the channel, resulting in the zero net charge
current.

In practical designs, the ``soft'' L contact might have different
properties from the remaining two. The exact left-right symmetry is
not required for the digital effect; any asymmetry between the
terminals can be counteracted by adjusting the voltages. Moreover,
the zero current state can be stabilized using the feedback loops
adjusting $V_{L}$ and $V_{R}$. Although the current from the L
contact is larger than the read-out R current, the L output needs
not be wasted. It can, for instance, be used to amplify the output
from the R contact using external circuits.

In all the modeling presented we have used experimentally available
properties of FM/SC structures. The formalism and ideas used here
can be applied as well to an all-metallic system, with a
paramagnetic metal channel. However, vastly different parameters
makes the control of the MR amplification effect more difficult. The
requirements of having a sizable AP electrochemical potential
splitting, of being able to control the voltages with desired
accuracy, and of keeping the current densities at reasonable levels,
are hard to reconcile. Thus, the semiconductor-based system is more
naturally suited for demonstration of voltage-controlled spin
transference.

In summary, we have constructed a room temperature theory of the
transference of the spin polarization in the current between a pair
of electrodes to another pair provided they are connected by spin
diffusion. The three-terminal structure effectively exploits the
spin accumulation in the semiconductor channel. One result, the
amplification of the magneto-resistive effect by voltage control,
may be of practical importance as the electrical read-out of
magnetic memory integrable to an electronics circuit. Such a  device
may be used in a scheme of ``magnetic computation'', working as a
building block of reprogrammable logic gate \cite{Ney_Nature03}.
Such a synergy of information processing and non-volatile storage
represents a possible fulfillment of the promises of spintronics.
Further properties which may result from the spin transference need
to be explored in the future.

\begin{acknowledgments}
This work is supported by NSF under Grant No. DMR-0325599.
\end{acknowledgments}


\begin{thebibliography}{99}
\bibitem{Baibich_PRL88}  M. N. Baibich \textit{et al.}, Phys.  Rev. Lett. \textbf{61}, 2472 (1988).
\bibitem{Binasch_PRB89} G. Binasch, P. Gr{\"u}nberg, F. Saurenbach, and W. Zinn, Phys. Rev.
B \textbf{39}, R4828 (1989).
\bibitem{Moodera_PRL95} J. S. Moodera, L. R. Kinder, T. M. Wong, and R. Meservey, Phys. Rev.
Lett. \textbf{74}, 3273 (1995).
\bibitem{Prinz_Science98} G. A. Prinz, Science \textbf{282}, 1660 (1998).
\bibitem{Wolf_Science00} S. A. Wolf \textit{et al.}, Science \textbf{294}, 1488 (2001).
\bibitem{Zutic_RMP04} I. {\u Z}uti{\'c}, J. Fabian, and S. D. Sarma, Rev. Mod.
Phys. \textbf{76}, 323 (2004).
\bibitem{Hanbicki_APL02} A. T. Hanbicki \textit{et al.}, Appl. Phys.
Lett. \textbf{80}, 1240 (2002).
\bibitem{Hanbicki_APL03} A. T. Hanbicki \textit{et al.}, Appl. Phys.
Lett. \textbf{82}, 4092 (2003).
\bibitem{Jiang_PRL05}  X. Jiang \textit{et al.}, Phys. Rev. Lett. \textbf{94}, 056601 (2005).
\bibitem{Adelmann_PRB05} C. Adelmann, X. Lou, J. Strand, C. J. Palmstrom, and P. A. Crowell, Phys. Rev. B \textbf{71}, 121301(R) (2005).
\bibitem{Fert_PRB01}   A. Fert and H. Jaffr{\`es}, Phys. Rev. B \textbf{64}, 184420 (2001).
\bibitem{Rashba_EPJB02} E. I. Rashba, Eur. Phys. J. B \textbf{29}, 513 (2002).
\bibitem{DattaDas_APL90}  S. Datta and B. Das, Appl. Phys. Lett \textbf{56}, 665 (1990).
\bibitem{Ciuti_APL02} C. Ciuti, J. P. McGuire, and L. J. Sham, Appl. Phys. Lett \textbf{81}, 4781 (2002).
\bibitem{Zutic_PRL02} I. {\u Z}uti{\'c}, J. Fabian and S. DasSarma, Phys. Rev. Lett. \textbf{88}, 066603 (2002).
\bibitem{Flatte_APL03} M. E. Flatt{\'e}, Z. G. Yu, E. Johnston-Halperin
and D. D. Awschalom, Appl. Phys. Lett \textbf{82}, 4740 (2003).
\bibitem{Schliemann_PRL03} J. Schliemann, J. C. Egues and D.
Loss, Phys. Rev. Lett. \textbf{90}, 146801 (2003).
\bibitem{Crooker_PRL05} S. A. Crooker and D. L. Smith, Phys. Rev. Lett. \textbf{94}, 236601 (2005).
\bibitem{Stephens_PRL04} J. Stephens, J. Berezovsky, J. P. McGuire, L. J. Sham, A. C. Gossard, D. D. Awschalom,   Phys. Rev. Lett. \textbf{93}, 097602 (2004).
\bibitem{Jedema_Nature01} F. J. Jedema, A. T. Filip, and B. J. van Wees, Nature \textbf{410}, 345
(2001).
\bibitem{Johnson_Science93} M. Johnson, Science \textbf{260}, 320
(1993).
\bibitem{Ney_Nature03} A. Ney, C. Pampuch, R. Koch, and K. H. Ploog, Nature \textbf{425}, 485
(2003).
\bibitem{Hershfield_PRB97} S. Hershfield and H. L. Zhao, Phys. Rev. B \textbf{56}, 3296
(1997).
\bibitem{Yu_Flatte_long_PRB02} Z. G. Yu and M. E. Flatt{\'e}, Phys. Rev. B \textbf{66}, 235302
(2002).
\bibitem{Rashba_PRB00}  E. I. Rashba, Phys. Rev. B \textbf{62}, R16267 (2000).
\bibitem{Albrecht_PRB03} J. D. Albrecht and D. L. Smith, Phys. Rev. B \textbf{68}, 035340
(2003).
\bibitem{tunneling_phenomena} \textit{Tunneling Phenomena}, edited by E. Burnstein and S. Lundqvist (Plenum Press, New York,
1969).
\bibitem{Butler_JAP97} W. H. Butler \textit{et al.}, J. Appl. Phys. \textbf{81}, 5518
(1997).
\bibitem{Optical_Orientation} \textit{Optical Orientation}, edited by F. Meier and B. P. Zakharchenya (Nort-Holland, New York,
1984).
\end{thebibliography}


 \end{document}